\documentclass[a4paper]{article}
\usepackage[dvips]{graphicx}
\usepackage{amsmath}
\usepackage{algorithm}
\usepackage{algorithmic}
\usepackage{url}

\usepackage{color}
\newcommand{\eref}[1]{(\ref{eq:#1})}
\newcommand{\elab}[1]{\label{eq:#1}}

\title{Morlet wavelet transform using attenuated sliding Fourier transform
  and kernel integral for graphic processing unit}

\author{Yukihiko~Yamashita
        \thanks{Yukihiko Yamashita is with the School of Environment and Society,
          Tokyo Institute of Technology, Tokyo, 152--8552, Japan
          email: yamasita@tse.ens.titech.ac.jp.}
        and~Toru~Wakahara
        \thanks{Toru Wakahara is with the Faculty of Computer and Information Sciences, 
          Hosei University, Tokyo, 184--8584, Japan,
          email: wakahara@hosei.ac.jp.}
        }
        
\begin{document}
\maketitle

\begin{abstract}
  Morlet or Gabor wavelet transforms
  as well as Gaussian smoothing, are widely used
in signal processing and image processing.
However, the computational complexity of their direct calculations
is proportional not only to the number of data points in a signal 
but also to the smoothing size, 
which is the standard deviation in the Gaussian function in their transform functions. 
Thus, when the standard deviation is large,
its considerable computation time diminishes the advantages
of aforementioned transforms.
Therefore, it is important to formulate an algorithm
to reduce the calculation time of the transformations. 
In this paper, we first review calculation methods of Gaussian smoothing by using
the sliding Fourier transform (SFT)
and our proposed attenuated SFT (ASFT) \cite{YamashitaICPR2020}.
Based on these methods, we propose two types of calculation methods for Morlet wavelet transforms. 
We also propose an algorithm to calculate SFT using the kernel integral on graphic processing unit (GPU).
When the number of calculation cores in GPU is not less than the number of data points,
the order of its calculation time is the logarithm of the smoothing size 
and does not depend on the number of data points.
Using experiments,
we compare the two methods for calculating the Morlet wavelet transform and
evaluate the calculation time of the proposed algorithm using a kernel integral on GPU.
For example, when the number of data points and the standard deviation are 102400 and 8192.0, respectively, 
the calculation time of the Morlet wavelet transform by the proposed method is
0.545 ms, which 413.6 times faster than a conventional method. 

\end{abstract}


\section{Introduction}

Morlet and Gabor wavelet transforms are widely used
in signal processing and image processing \cite{morletwavelet1984,LIN2000135}.
The calculation time by their direct calculation methods 
is proportional not only to the data size of a signal 
but also to the smoothing size, 
which is the standard deviation in the Gaussian function in their transform functions. 
In order to reduce them, 
many calculation methods such as an algorithm using the block processing,
the FFT, and two-point convolution algorithm for DSP
\cite{FastConv1999, Compo1998,Cao2014,Cohen2019,Yi2019} have been proposed.
However, their calculation times are proportional to not less than data size.
In this paper, we solve this problem by an algorithm that is suitable for parallel computing
of which calculation time does not depend on data size
when the number of processors is larger than the data size.

Gaussian smoothing has a similar formula as the Morlet wavelet transform
and is also very important for signal processing and image processing. 
Efficient and practical calculation methods of Gaussian smoothing including its first and second differentials
have been intensively researched; it has also been applied to image processing and computer vision
\cite{SIFTLowe2004,PCISIFTKe2004,Bay06surf:speeded,Bay:2008:SRF:1370312.1370556,WakaharaPAMI2001,YamashitaPR2016,ZhangIPTA2017,Iijima1989,ScaleSpaceFiltering1983}. 
In order to calculate Gaussian smoothing efficiently with a large smoothing size,
several methods have been proposed such as the fast Fourier transform (FFT) \cite{FFT1965},
recursive methods \cite{Deriche1990,Farneback2006},
and methods based on the sliding Fourier transform (SFT) \cite{SlideFourierKober2004}.
Essentially, the computational complexities of SFT by a recursive filter and FFT do not depend on the smoothing size.
In particular, methods based on SFT are computationally most efficient.
Their precision can be controlled easily.
Moreover, their algorithm is simple and suitable for parallel processing.
Hence, in this paper, we focus on the methods based on SFT
to accelerate the calculation of the Morlet wavelet transform.

We introduce the calculation methods of Gaussian smoothing by SFT. 
Elboher et al.~\cite{Elboher2011CosineII} proposed using the kernel integral as SFT to calculate Gaussian smoothing. 
Sugimoto et al.~\cite{SugimotoICASSP2018} conducted an in-depth study on calculating Gaussian smoothing by the sliding discrete cosine transform (DCT).
They applied various types of sliding DCTs calculated by recursive filters,
and compared their precisions and computational efficiencies.
Moreover, the same research group \cite{SugimotoDCT-1-2018} recommended using the sliding DCT-1
of which basic wavelength is the twice of the data length.
Yamashita et al.~\cite{Yamashita2020} used 
Gaussian smoothing and its differentials by SFT for object detection in a scene.
Moreover, they \cite{YamashitaICPR2020} highlighted the problem of lack of precision
when using the single-precision floating-point calculation.
Because SFT is given by the difference between the integrated values at two points and integrated values likely diverges, 
the precision of the difference becomes low especially when using the single-precision floating-point calculation.
To solve the aforementioned problem,
they proposed the attenuated SFT (ASFT) \cite{YamashitaICPR2020}.

In this paper, we first review calculation methods of Gaussian smoothing using the SFT/ASFT.
Based on these methods, we propose two methods for calculating the Morlet wavelet transform. 
The first one approximates the Morlet wavelet transform directly by SFT/ASFT.
The second method is realized by multiplying a Gaussian function approximated by SFT/ASFT and sinusoidal functions.
Because both functions are expressed by sinusoidal functions, their product are also expressed by sinusoidal functions; 
hence the transform can be approximated by SFT/ASFT.
We also propose an algorithm to calculate SFT using the kernel integral on graphic processing unit (GPU) \cite{CUDA2018,ParallelReduction201X}.
When the number of calculation cores in GPU is not less than the number of data points,
the order of its calculation time is the logarithm of the smoothing size 
and does not depend on the number of data points in a signal.
We finally perform comparison and evaluation. 
The purpose of the experiments is twofold.
The first purpose is to compare the two methods for calculating the Morlet wavelet transform.
The second purpose is to evaluate the computation time of the proposed algorithm using the kernel integral on GPU. 

The remainder of this paper is organized as follows.
In Section II, we summarize the calculation methods of Gaussian smoothing using the SFT/ASFT.
In Section III, we propose two methods for calculating the Morlet wavelet transform.
In Section IV, we propose an algorithm to calculate Gaussian smoothing and the Morlet wavelet transform on GPU.
In Section V, we perform experiments to compare the two methods for the Morlet wavelet transform and
show the advantages of the proposed algorithm for Gaussian smoothing and the Morlet wavelet transform on GPU.
Section VI concludes this paper.  

\section{Gaussian smoothing and its differentials using SFT/ASFT}

For a standard deviation $\sigma$, let $\gamma = 1 / (2 \sigma ^2)$.
Then, the Gaussian function is defined by 
\begin{equation}
G[n] =\sqrt{\frac{\gamma }{\pi}} e^{-\gamma n^2}. 
\end{equation}
We also define the differentials of the Gaussian function as
\begin{align}
G_{\rm D}[n] & \equiv  (-2 \gamma n) G[n], \\
G_{\rm DD}[n] & \equiv  (4\gamma^2 n^2 -2 \gamma) G[n]
\end{align}

For an input signal $x[n]$ defined in $[0, N-1]$ with an integer $N$,
Gaussian smoothing and its first and second differential signals are given respectively by
\begin{align}
x_{\rm G}[n] & \equiv  \sum_{k=-K}^K G[k] x[n-k], \elab{xG}\\
x_{\rm GD}[n] & \equiv  \sum_{k=-K}^K G_{\rm D}[k] x[n-k], \\
x_{\rm GDD}[n] & \equiv  \sum_{k=-K}^K G_{\rm DD}[k] x[n-k]. \elab{xGDD}
\end{align}
For the calculation, we assume that values of $x[n]$ are extended properly 
even if $n < 0$ or $n \geq N$.
Usually, they are either zero or the values on the edges of the interval.
Because $G[n]$ decays rapidly,  
the summation is calculated in an interval $[-K,K]$ for an integer $K$.
The complexity of the three calculations of
eqs.\eref{xG}--\eref{xGDD} is $O(KN)$.
A larger $\sigma$ requires larger $K$ and more calculation steps.

\subsection{Gaussian smoothing and its differentials using SFT}

Let $\beta = \pi /K$.
Let $c_p[n]$ and $s_p[n]$ be the $p$-th order components of SFT 
of interval $[-K,K]$ for the input signal $x[n]$. They are given by
\begin{align}
c_p[n] & = \sum_{k=-K}^K x[n-k] \cos (\beta p k), \elab{cp}\\
s_p[n] & = \sum_{k=-K}^K x[n-k] \sin (\beta p k). \elab{sp}
\end{align}
Note that the bases in this SFT are not orthogonal.

We define approximations of the Gaussian function ($-K \leq k \leq K$) and its differentials with Fourier series by 
\begin{align}
\hat{G}[k]          & \equiv  \sum_{p=0}^P a_p \cos (\beta p k) \simeq G[k], \elab{aproGauss}\\
\hat{G}_{\rm D}[k]   & \equiv  \sum_{p=1}^P b_p \sin (\beta p k) \simeq G_{\rm D}[k], \\
\hat{G}_{\rm DD}[k]  & \equiv  \sum_{p=0}^P d_p \cos (\beta p k) \simeq G_{\rm DD}[k].
\end{align}
Here, we denote $G$, $G_{\rm D}$, $G_{\rm DD}$ simply by $G_{\rm X}$ for brevity.
Then, each set of coefficients $a_p$, $b_p$, and $d_p$ can be decided by
the minimum mean square error (MMSE), that is, minimizing 
\begin{equation}
 \sum_{k=-K}^K \left| \hat{G}_{\rm X}[k] - G_{\rm X}[k]   \right|^2, \elab{mmse}  
\end{equation}

The smoothed signals and their differentials are given by 
\begin{align}
x_{\rm G}[n] & \simeq \sum_{p=0}^P a_p c_p[n] \elab{GsSFT},\\
x_{\rm GD}[n] & \simeq  \sum_{p=1}^P b_p s_p[n],\\
x_{\rm GDD}[n] & \simeq  \sum_{p=0}^P d_p  c_p[n].
\end{align}
As we discuss in the next subsection,
each component of the SFT can be calculated with a computational complexity
of $O(N)$, whereas the complexity of smoothing is the order of $O(PN)$.
By setting the value of $P$ from 2 to 6, we can substantially decrease the calculation steps as compared with
calculating the convolution between the signal and the truncated Gaussian function.

Values with sufficient precision have been obtained at $P=4$ when calculating smoothed differentials for an image 
and $P=2$ when smoothing edge direction histograms
are used for object detection \cite{Yamashita2020}.

\subsection{SFT using kernel integrals}

We can simplify the method for calculating the SFT with 
complex expressions of Fourier transform.
We define the kernel integral of an input signal $x[n]$ by 
\begin{equation}
  \elab{kernel}
  u[n] = \sum_{k=0}^n x[k] e^{i \beta p k}.
\end{equation}
The values can be calculated by the following recurrence formula:
\begin{equation}
  \elab{kernelZenka}
u[n] = u[n-1] + x[n] e^{i \beta p n}  .
\end{equation}
The kernel integral of interval $[n-2K, n]$ is defined by 
\begin{equation}
  \elab{kernelInt}
  u_{(2K+1)}[n] = \sum_{k=n - 2K}^{n} x[k] e^{i \beta p k},
\end{equation}
and calculated as
\begin{equation}
  \elab{kernelInter}
u_{(2K+1)}[n] = u[n] - u[n - 2K - 1]. 
\end{equation}
Then, the SFT is given by 
\begin{equation}
c_p[n] - i s_p[n] = e^{-i \beta p n} u_{(2K+1)}[n+K].
\end{equation}
According to the aforementioned relation,
SFT can be calculated with a computational complexity of $O(N)$ for each $p$.

We can also calculate $u_{(2K+1)}[n]$ directly by the following recurrence formula.
\begin{align}
  u_{(2K+1)}[n] =  u_{(2K+1)}[n-1] + x[n] e^{i \beta p n} 
                   - x[k - 2K - 1] e^{i \beta p (n - 2K - 1)}. \elab{kernelIntZenka}
\end{align}

\subsection{SFT using recursive filters}

In this subsection, we explain the calculation of SFT
using recursive filters.
First, we define a first-order recursive filter by  
\begin{equation}
v[n] = e^{-i \beta p} v[n-1] + x[n]. \elab{iir1}
\end{equation}
The output $v[n]$ of this filter can be considered as the summation
\begin{equation}
v[n] = \sum_{k}^n e^{-i \beta pk} x[n-k].
\end{equation}
We define values when the interval of the summation is set at $2K+1$ and $2K$ for $v_{(2K+1)}$ and $v_{(2K)}$, respectively. As a result, they can be given by
\begin{align}
v_{(2K+1)}[n]  & =  v[n] - e^{-i \beta p} v[n - 2K - 1], \elab{iir12K1}\\
v_{(2K)}[n]    & =  v[n] - v[n - 2K]. \elab{iir12K}
\end{align}
Because we have $e^{\beta 2K} = 1$, eq.\eref{iir12K} is simpler than eq.\eref{iir12K1}.

From either of the values, the results of SFT can be given by
\begin{align}
  c_p[n] - i s_p[n] & =  (-1)^p v_{(2K+1)}[n + K], \elab{sfiir12K1}\\
  c_p[n] - i s_p[n] & = (-1)^p (v_{(2K)}[n + K] + x[n - K])  . \elab{sfiir12K}
\end{align}
Because four multiplications of real values are necessary for calculating
a multiplication of two complex values,
it is faster to use the truncation by $2K$ (eqs. \eref{iir12K} and \eref{sfiir12K}) than to use the truncation by $2K + 1$ (eqs. \eref{iir12K1} and \eref{sfiir12K1}).

\begin{figure}
  \centering
  \includegraphics[keepaspectratio,width=7.4cm]{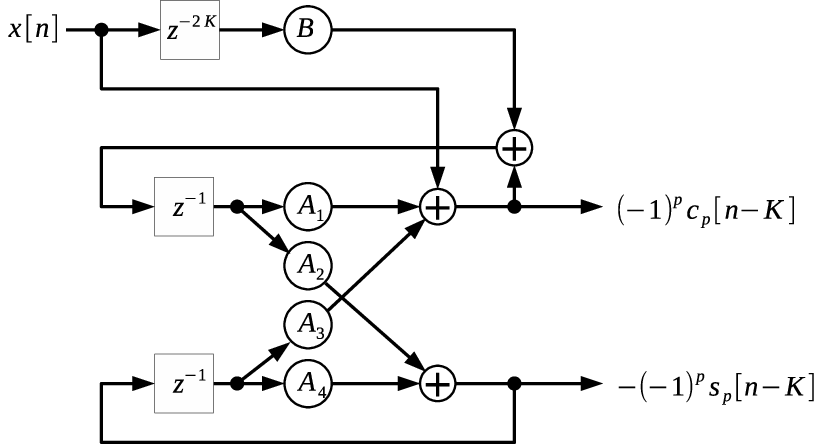}

  \
  
  \caption{Block diagram of recursive FIR filter to calculate SFT.}
  \label{fig:blockFIR}
\end{figure}

$v_{(2K)}[n]$ can be directly calculated by the following recurrence formula.
\begin{equation}
v_{(2K)}[n] =  e^{-i \beta p} v_{(2K)}[n-1] + x[n] - x[n -2K].  \elab{iir1IntZenka}
\end{equation}
Fig.~\ref{fig:blockFIR} illustrates the block diagram of the recursive FIR (finite impulse response) filter 
\eref{iir1IntZenka} where $A_1 = A_4 = \cos \beta p$, $-A_2 = A_3 = \sin \beta p$, 
and $B=-1$.

Sugimoto et al.~proposed to use second-order recursive filters. 
When we substitute $n-1$ for $n$ in eq.\eref{iir1}, we have
\begin{equation}
v[n-1] = e^{-i \beta p}v[n-2] + x[n-1]. \elab{iir2m}  
\end{equation}
Multiplying eq.\eref{iir2m} by $e^{i \beta p}$ and subtracting it from eq.\eref{iir1},
we have a second-order recursive filter as 
\begin{equation}
v[n] = 2 \cos (\beta p) v[n-1] - v[n-2] + x[n] - e^{i \beta p} x[n-1]. \elab{iir2Zenka}
\end{equation}
Because $v[n]$ is multiplied by only real values,  
we can separate the real and imaginary values
to calculate $c_p[n]$ and $s_p[n]$, respectively.
However, as the formula is similar to the second-order differential equation, 
it might result in a large calculation error by floating-point operations.
Note that the values of $v[n]$ in the first- and second-order recursive filters are the same
while containing different calculation errors.

$v_{(2K)}[n]$ can be directly calculated by the following recurrence formula
corresponding to the second-order recursive filter.
\begin{align}
v_{(2K)}[n]  = 2 \cos (\beta p) v_{(2K)}[n-1] - v_{(2K)}[n-2] 
                + x[n] - x[n-2K]
                 - e^{i \beta p}(x[n - 1] - x[n - 2K - 1]) \elab{iir2IntZenka}
\end{align}

\subsection{Gaussian smoothing and its differentials using ASFT}

Nowadays, the application of GPU is needed to shorten computation times for signal processing problems. 
However, economical GPUs only have single-precision floating-point units.
Because the values of filters such as $u[n]$ and $v[n]$ accumulate errors
as $n$ increases,
it causes a large error in Gaussian smoothing,
especially when we use single-precision floating-point units.
In order to reduce the accumulated errors in the values of filters,
Gaussian smoothing by ASFT was proposed. 

Let $\alpha (>0)$ and $\beta$ be a decay constant and $\pi /K $, respectively.
And let $\tilde{c}_p[n]$ and $\tilde{s}_p[n]$ be the $p$-th order components of ASFT 
of interval $[-K,K]$ for the input signal $x[n]$. They are given by
\begin{align}
\tilde{c}_p[n] & =  \sum_{k=-K}^K x[n-k] e^{\alpha k} \cos (\beta p k), \elab{acp}\\
\tilde{s}_p[n] & =  \sum_{k=-K}^K x[n-k] e^{\alpha k} \sin (\beta p k). \elab{asp}
\end{align}
The weights for older data $(k<0)$ become small exponentially.  

We define a first-order recursive filter by
\begin{equation}
\tilde{v}[n] = e^{-\alpha -i \beta p} \tilde{v}[n-1] + x[n]. \elab{iir1A}
\end{equation}
We define its filter values of interval of length $2K$ as
\begin{equation}
\tilde{v}_{(2K)}[n]  = \tilde{v}[n] - e^{- \alpha (2K) } \tilde{v} [n - 2K]. \elab{aiir12K}
\end{equation}
Then, the ASFT can be calculated by
\begin{equation}
   \tilde{c}_p[n] - i \tilde{s}_p[n]
  =  (-1)^p e^{-\alpha K} (\tilde{v}_{(2K)}[n + K] + e^{- 2 \alpha K} x[n - K])).  \elab{asfiir12K}
\end{equation}

When $\alpha > 0$,
by the attenuation factor $e^{-\alpha} $ in eq.\eref{iir1A}, $\tilde{v}[n]$ is bounded for bounded input $x[n]$. 
As a result, the calculation of the ASFT is stabilized
even using single-precision floating-point calculations.

$\tilde{v}_{(2K)}[n]$ can be directly calculated by the following recurrence formula.
\begin{equation}
\tilde{v}_{(2K)}[n] = e^{- \alpha -i \beta p} \tilde{v}_{(2K)}[n-1] + x[n] - e^{- 2\alpha K}x[n -2K] 
\end{equation}

$\tilde{v}[n]$ can be calculated by a second-order recursive filter as
\begin{align}
\tilde{v}[n] =  2 e^{-\alpha} \cos (\beta p) \tilde{v}[n-1] - e^{-2 \alpha} \tilde{v}[n-2] 
   + x[n] - e^{-\alpha + i \beta p} x[n-1]. \elab{iir2AZenka}
\end{align}
Moreover, $\tilde{v}_{(2K)}[n]$ can be calculated by
\begin{align}
\tilde{v}_{(2K)}[n] & = 2 e^{- \alpha  }\cos (\beta p) \tilde{v}_{(2K)}[n-1] - e^{-2 \alpha  }\tilde{v}_{(2K)}[n-2] 
                      + x[n] -e^{-2 \alpha K } x[n-2K] \nonumber \\
                     & - e^{- \alpha  }e^{i \beta p}(x[n - 1] - e^{-2 \alpha K } x[n - 2K - 1]). \elab{iir2AIntZenka}
\end{align}

\subsection{Gaussian smoothing and its differentials using ASFT}

If we directly use ASFT instead of SFT, it causes an error.
However, in the case of a Gaussian function, because we have 
\begin{equation}
  e^{\alpha n} e^{-\gamma n^2}  = e^{ \frac{\alpha^2}{4\gamma}} e^{-\gamma \left(n - \frac{\alpha}{2\gamma}\right)^2},
  \elab{siftGauss}
\end{equation}
the attenuation can be converted to an translation in the time domain.
By selecting an integer as the amount of shift to an integer and shifting the output,
we can obtain the true Gaussian smoothed signal and its differentials.

Let we choose $\alpha$ such that $n_0 = \frac{\alpha}{2\gamma}$ becomes an integer.
It follows that
\[
G[n - n_0] = e^{- \frac{\alpha^2}{4\gamma}} e^{\alpha n} G[n].
\]

Eq.\eref{aproGauss} yields
\begin{equation}
  \sum_{k=-K}^K G[k] \left(e^{\alpha k} x[n-k] \right) \simeq \sum_{p=0}^P a_p \tilde{c}_p[n].
\end{equation}
Then, we have
\begin{align}
\tilde{x}_{\rm G}[n] \equiv  \sum_{k=-K}^K G[k-n_0] x[n-k] 
 =  e^{- \frac{\alpha^2}{4\gamma}} \sum_{k=-K}^K G[k] \left(e^{\alpha k} x[n-k] \right) 
\simeq   e^{ -\frac{\alpha^2}{4\gamma}} \sum_{p=0}^P a_p \tilde{c}_p[n].
\end{align}

Moreover, we have 
\begin{equation}
  \tilde{x}_{\rm G}[n]
  =  \sum_{k=-K-n_0}^{K-n_0} G[k] x[(n - n_0) - k]. \elab{asftcal5}
\end{equation}
We assume that $n_0$ is small compared to $\sigma$.
Because the values of $G[n]$ for $|n| \geq \sigma$ decrease rapidly,
we have
\begin{align}
 \sum_{k=-K-n_0}^{-K-1} G[k] x[(n - n_0) - k] \simeq & 0, \nonumber \\
 \sum_{k=K-n_0 + 1}^{K} G[k] x[(n - n_0) - k] \simeq & 0. \nonumber
\end{align}
Hence, eq.\eref{asftcal5} yields
\begin{equation}
  \tilde{x}_{\rm G}[n] \simeq x_{\rm G}[n-n_0]. 
\end{equation}
Finally, we have
\begin{equation}
x_{\rm G}[n] \simeq  e^{ -\frac{\alpha^2}{4\gamma}} \sum_{p=0}^P a_p \tilde{c}_p[n + n_0]. \elab{GsASFT}
\end{equation}

Although in the case of differentials we have to consider the polynomial factors
in addition to the Gaussian function, we have the following relation.  
\begin{align}
x_{\rm GD}[n] \simeq & e^{ -\frac{\alpha^2}{4\gamma}} \sum_{p=0}^P \alpha a_p \tilde{c}_p[n + n_0]  - \sum_{p=1}^P  2\gamma b_p \tilde{s}_p[n + n_0], \elab{gddAsft}\\
  x_{\rm GDD}[n] \simeq & e^{ -\frac{\alpha^2}{4\gamma}} \sum_{p=0}^P (d_p + \alpha^2 a_p ) \tilde{c}_p[n + n_0] 
                + \sum_{p=1}^P (2 \alpha)\tilde{b}_p s_p[n + n_0] .
\end{align}
To deal with the polynomial factors, it is necessary
to use both $\cos$ and $\sin$ components 
to calculate the first and second differentials of Gaussian smoothing, respectively.
Although the calculation time of ASFT might be longer than that of SFT,
experiments \cite{YamashitaICPR2020} showed that their differences are small. 

\begin{table}[t]
  \caption{Relative RMSE of Gaussian smoothing and its differentials with coefficients by MMSE
  ($K=256$ and $n_0 = 10$) \cite{YamashitaICPR2020}.}
\begin{center}
\begin{tabular}{|l|c|r|r|r|}
  \hline
Transform  & P & $e(G)$ (\%)& $e(G_{\rm D})$  (\%)& $e(G_{\rm DD})$ (\%) \\
  \hline
    & 2 & 1.0 & 5.1 & 8.2 \\
  \cline{2-5}
    & 3 & 0.15 &  0.90 &  2.77  \\
  \cline{2-5}
SFT & 4 & 0.038 & 0.24 & 0.54  \\
  \cline{2-5}
    & 5 & 0.0059 & 0.043 & 0.16  \\
  \cline{2-5}
    & 6 & 0.0015 & 0.011 & 0.031  \\
  \hline
    & 2 & 1.1 & 5.4 & 8.5 \\
  \cline{2-5}
    & 3 & 0.17 & 1.02 & 3.10 \\
  \cline{2-5}
ASFT & 4 & 0.046 & 0.30 & 0.63 \\
  \cline{2-5}
    & 5 & 0.017 & 0.037 & 0.12 \\
  \cline{2-5}
    & 6 & 0.0021 & 0.016 & 0.041 \\
  \hline
\end{tabular}
\label{tab1mse}
\end{center}
\end{table}

Table \ref{tab1mse} shows the following relative root mean square errors (relative RMSEs) $e(G_{\rm X})$ using SFT and ASFT.
\begin{equation}
 e(G_{\rm X}) = \sqrt{\frac{\sum_{n=-3K}^{3K} \left| \hat{G}_{\rm x}[n] - G_{\rm x}[n]  \right|^2}{\sum_{n=-3K}^{3K} \left| G_{\rm x}[n]  \right|^2}}. \elab{mmseGEx}
\end{equation}
Here, we can neglect errors by floating-point 
because we use the double-precision method.
We let values of the approximated Gaussian function and its differentials outside $[-K, K]$ be 0.
Because $K$ is close to $3\sigma$ ($3K \simeq 9 \sigma$) in eq.\eref{mmseGEx},
the interval $[-3K,\,3K]$ is sufficient to evaluate the precision of the Gaussian function. 
$K$ is fixed at 256 and the parameter $\beta$ for each $P$ is decided
as relative RMSEs are minimized.
For differentials of Gaussian smoothing, the same $\beta$ is used for each $P$.
For ASFT, $\alpha$ is decided such that $n_0 = 10$. 

From Table \ref{tab1mse} we can observe that $P=3$ has sufficient precision for SFT and ASFT. 
This is because the relative RMSE of a Gaussian function is 0.46\% 
after truncating within the interval of $3 \sigma$ that is typically used for many applications.

\section{Morlet wavelet transform via SFT/ASFT}
The Morlet wavelet function is defined by
\begin{equation}
\psi_{\xi}(t) = \frac{C_{\xi}}{\pi ^{\frac{1}{4}}} e^{-\frac{1}{2}t^2} (e^{i \xi t} - \kappa_{\xi}),
\end{equation}
where
\begin{align}
C_{\xi} & = \left(1 + e^{-\xi^2} - 2 e^{-\frac{3}{4}\xi^2} \right)^{-\frac{1}{2}},\\
\kappa_{\xi} & = e^{-\frac{1}{2} \xi ^2}.
\end{align}
This formula is based on a Gaussian function for which the standard deviation is 1.0.
In many applications, this function is dilated by a factor $\sigma$ that is analogous to the standard deviation of Gaussian functions and is used as a discrete function as:
\begin{equation}
\psi_{\sigma, \,\xi}[n] = \frac{c_{\xi}}{\pi ^{-\frac{1}{4}} \sqrt{\sigma}} e^{-\frac{1}{2 \sigma^2 }n^2} (e^{i \frac{\xi}{\sigma}  t} - \kappa_{\xi}).
\end{equation}

Approximation of $\psi_{\sigma, \,\xi}[n]$ by the sum of sinusoidal functions provides
a method for calculating the Morlet wavelet transform using SFT/ASFT.
To approximate $\psi_{\sigma, \,\xi}[n]$, we can basically consider two methods.
The first one is to approximate it directly by the sum of 
sinusoidal functions of orders from $P_{\rm S}$ to $ P_{\rm S} + P_{\rm D} - 1$.
The second one is to approximate a Gaussian function by the sum of sinusoidal functions
as shown in Section II, and to multiply it with $(e^{i \frac{\xi}{\sigma} t} - \kappa_{\xi})$.
Because the products are the sum of sinusoidal functions,
the Morlet wavelet transform can be calculated by SFT/ASFT.
We call these two methods, the direct method and multiplication method, respectively.
\subsection{Direct method}

The direct method approximates $\psi_{\sigma, \,\xi}[k]$ ($-K \leq k \leq K$) by
\begin{equation}
\psi_{\sigma, \,\xi}[k]  \simeq  \sum_{P_{\rm S}}^{P_{\rm S} + P_{\rm D} - 1}
                            (m_p \cos (\beta p k) + i l_p \sin(\beta p k)).
\end{equation}
The coefficients $m_p$ and $l_p$ ($P_{\rm S} \leq p \leq P_{\rm S} + P_{\rm D} - 1$) are calculated by MMSE.
Then, the result of the Morlet wavelet transform, $x_{\rm M}[n]$ is given by
\begin{equation}
  x_{\rm M}[n]  \simeq \sum_{p=P_{\rm S}}^{P_{\rm S} + P_{\rm D} - 1} (m_p c_p[n] + i l_p s[n]).
\end{equation}
In this method, almost the same algorithm \eref{GsSFT} and \eref{GsASFT} for Gaussian smoothing can be used.
When we use ASFT, the following formula can be obtained. 
\begin{equation}
  x_{\rm M}[n] \simeq  e^{ -\frac{\alpha^2}{4\gamma}} 
  \sum_{p=P_{\rm S}}^{P_{\rm S} + P_{\rm D} - 1} (m_p \tilde{c}_p[n + n_0] + i l_p \tilde{s}_p[n + n_0]).
\end{equation}

\subsection{Multiplication method}
We define $a'_p$ by
\begin{equation}
  a'_p =
  \left\{
    \begin{array}{ll}
      \frac{1}{2} a_p & (p>0)\\
       a_p & (p=0)\\
      \frac{1}{2} a_{-p} & (p<0)\\
  \end{array}
    \right..
  \end{equation}
  Then, we have 
\begin{equation}
G[n] \simeq  \sum_{p=0}^P a_p \cos (\beta p n) = \sum_{p=-P}^P a'_p e^{i \beta p n}.
\end{equation}
We expand SFT when the frequency is a real number. 
\begin{align}
c(\omega)[n] & = \sum_{k=-K}^K x[n-k] \cos (\omega k), \\
s(\omega)[n] & = \sum_{k=-K}^K x[n-k] \sin (\omega k). 
\end{align}

Let $\omega_p = \frac{\xi}{\sigma} + \beta p$. Then we have
\begin{align}
  x_{\rm M}[n] \simeq  \frac{C_{\xi}}{\pi ^{\frac{1}{4}} \sqrt{\sigma}} 
                        \sum_{p=-P}^P \left (a'_p c(\omega_p)[n] + i a'_p s(\omega_p)[n] \right. 
   \left. + \kappa_{\xi} a_p c_p[n]\right).
\end{align}

When using ASFT, we can obtain the following equation.
\begin{align}
  x_{\rm M}[n] \simeq \frac{C_{\xi} e^{-\frac{\sigma^2}{4 \gamma}}}{\pi ^{\frac{1}{4}} \sqrt{\sigma}}  \sum_{p=-P}^P  \left (a'_p \tilde{c}(\omega_p)[n + n_0]  \right. 
 \left. + i a'_p \tilde{s}(\omega_p)[n + n_0]  + \kappa_{\xi} a_p \tilde{c}_p[n + n_0]\right), 
\end{align}
where $\tilde{c}(\omega)$ and $\tilde{s}(\omega)$ can be defined similarly as in eqs.\eref{acp} and \eref{asp}.

In the multiplication method, we have to use an SFT
whose frequency does not match with $K$.
That is, $\pi/\xi$ may not be a multiplier of $K$.
However, it can be calculated by the kernel integral or recursive filters.
However, the processes to align the center and make the length of the summation $2K + 1$ are complicated.
Therefore, if the multiplication method does not have an advantage in precision,
it is better to use the direct method.

\section{GPU implementation of SFT using kernel integral}

The order of calculation complexity of the SFT/ASFT transformation is $O(PN)$,
where $P$ is $P_{\rm D}$ or $P_{\rm M}$.
It does not depend on $K$ corresponding to the standard deviation $\sigma$.
Let us consider the case where we use recursive filters calculated on GPU. 
When an image of size $N_{\rm X} \times N_{\rm Y}$ is filtered,
lines in the image are independently calculated; 
hence calculation time is $O(P(N_{\rm X}+N_{\rm Y}))$.
The method has an advantage in calculation for two-dimensional (2D) signals
because the number of calculation cores in a current GPU
is larger than the number of lines and smaller than the number of pixels. 

In this section, we propose a new algorithm of SFT on GPU using the kernel integral for one-dimensional (1D) signals.
When the number of calculation cores is larger than data size $N$, 
its calculation time is $O(P \log_2 K)$, that is,
the calculation time does not depend on the data size.

When we use a GPU, the most time-consuming calculation of the SFT by the kernel integral is 
the summation of $x[k] e^{i \beta p k}$ in the interval $[n-2K, n]$ for each $n$ in eq.\eref{kernelInt}
because the other calculations can be done point-wise.
We call this calculation the sliding sum.

From here, we calculate a sliding sum of a signal $f[n]$ as
\begin{equation}
h[n] = \sum_{k = 0}^{L-1} f[n + k] 
\end{equation}
for $n = 0,\,1,\,\ldots,\,N - L$, where $L$ is the length of the summation.

\subsection{Basic algorithm for the sliding sum}

At first, we do not consider shared memory in GPU
and explain an algorithm that only uses global memory.
Let $R$ be an integer such that $2^{R - 1} \leq L < 2^R$.
We define a functions $B(m,\, r)$ for integers $m$ and $r$ by
\begin{align}
B(m,\,r) \equiv \lfloor m / 2^{r} \rfloor \,\,\, \mbox{mod 2}.
\end{align}
$B(m,\,r)$ is the same as the value of the $r$th bit with the binary expression of $m$
($0$-th bit is the least significant bit).

At first, we let $g_0[n] = f[n]$ and $h_0[n] = 0$.
For $r = 0,\, 1,\,\ldots,\, R - 1$ and $n = 0,\, 1,\,\ldots,\, N - 2^r -1$, we calculate
\begin{align}
g_{r+1} [n]& = g_{r}[n] + g_{r}[n + 2^r]. \\
h_{r+1} [n]& = \left\{
  \begin{array}{ll}
  g_{r}[n] + h_{r}[n + 2^r]  & \mbox{if $B(L,\,r) = 1$},\\
  h_{r}[n].  & \mbox{otherwise} \\
  \end{array} \right. .
\end{align}
Then, $h[n]$ is given by $h_{R}[n]$. This algorithm is summarized in Algorithm \ref{al:ssBasic}. 

\begin{algorithm}
\caption{Basic algorithm for the sliding sum.} 
\label{al:ssBasic}
\begin{algorithmic}
\REQUIRE $f[\,], \,L$
\STATE Obtain $l$ such that $2^{R -1} \leq L < 2^R$.
\STATE $g[n] \gets f[n]$ for $n = 0,\,\ldots,\, N-1$.
\STATE $h[n] \gets 0.0$ for $n = 0,\,\ldots,\, N-1$.
\FOR{$r \in \{0,\,1,\,...,\,R-1\}$}
\IF{$B(L,\,r) = 1$}
\STATE $h[n] \gets g[n] + h[n + 2^r]$ for $n = 0,\,..., N - 1 - 2^r$
\ENDIF
\STATE $g[n] \gets g[n] + g[n + 2^r]$ for $n = 0,\,..., N - 1 - 2^r$
\ENDFOR
\RETURN $h[\,]$
\end{algorithmic}
\end{algorithm}

\subsection{Algorithm for the sliding sum with shared memory}

Because the speed of global memory of GPU is slower than that of shared memory,
usage of shared memory in a block is very effective.
Here, we assume that the sliding sum of eight elements is done in a block,
although the sum of more elements is possible by a recent GPU.

After processing each block, the data have to be arranged for the next stage
to sum up more than eight elements as shown in Figure \ref{fig:arrange}. 
By arranging and adding the next elements in the horizontal direction,
we can easily calculate the sums of elements of distances of 8, 16, and 32 in the original order.

\noindent
\begin{figure*}
  \centering
  
\begin{center}
\begin{tabular}{|p{2mm}|p{2mm}|p{2mm}|p{2mm}|p{2mm}|p{2mm}|p{2mm}|p{2mm}|p{2mm}|p{2mm}|p{2mm}|p{2mm}|p{2mm}|p{2mm}|p{2mm}|p{2mm}|p{2mm}|p{2mm}|p{2mm}|p{2mm}|p{2mm}|p{2mm}|p{2mm}|p{2mm}|p{2mm}|p{2mm}|p{2mm}|}
\hline
0 & 1 & 2 & 3 & 4 & 5 & 6 & 7 & 8 & 9 & 10 & 11 & 12 & 13 & 14 & 15 & 16 & 17 & 18 & 19 & 20 & 21 & 22 & 23 & $\cdots$ \\
\hline
\end{tabular}
\end{center}
(a) Before arrangement

\

\begin{center}
\begin{tabular}{|p{2mm}|p{2mm}|p{2mm}|p{2mm}|}
\hline
0 & 8 & 16 & $\cdots$ \\
\hline
1 & 9 & 17 & $\cdots$ \\
\hline
2 & 10 & 18 & $\cdots$ \\
\hline
3 & 11 & 19 & $\cdots$ \\
\hline
4 & 12 & 20 & $\cdots$ \\
\hline
5 & 13 & 21 & $\cdots$ \\
\hline
6 & 14 & 22 & $\cdots$ \\
\hline
7 & 15 & 23 & $\cdots$ \\
\hline
\end{tabular}
\end{center}
(b) After arrangement

\caption{Rearrangement of elements.}
  \label{fig:arrange} 
\end{figure*}

We define the following notations.
\begin{itemize}
\item $x_{\rm T}$: the thread id in $x$-coordinate. (Given in a thread.)
\item $x_{\rm B}$: the block id in $x$-coordinate. (Given in a thread.)
\item $y_{\rm T}$: the thread id in $y$-coordinate. (Given in a thread.)
\item $y_{\rm B}$: the block id in $y$-coordinate. (Given in a thread.)
\item $N_8 = 8^{x}$: The minimum integer among those satisfying $8^{x} \geq N$.
\item $g_1[\,][\,]$,\, $g_2[\,][\,]$,\, $h_1[\,][\,]$,\, $h_2[\,][\,]$: 2D arrays in global memory of which the total length is $N_8$. 
  (Although the shape of the arrays at different stages are different, the total length is the same.) 
\item $s[\,][\,]$,\, $t[\,][\,]$: 2D array in shared memory in a block of size $(16, 8)$. 
\end{itemize}

\begin{algorithm}
  \caption{Program for the sliding sum in CPU.}
  \label{al:ssCPU}
\begin{algorithmic}[1]
  \REQUIRE $f[\,]$, $N$, $L$
  \STATE Clear all values in $g_1[\,][\,]$,\, $g_2[\,][\,]$,\, $h_1[\,][\,]$,\, and $h_2[\,][\,]$ to 0.0. 
\STATE $g_1[n][0] \gets f[n]$ for $n = 0,\,...,\, N-1$.
\WHILE{$L > 0$}
\STATE ($g_2[\,][\,],\, h_2[\,][\,],\, L) \gets \textsc{SSSG}(g_1[\,][\,],\, h_1[\,][\,],\, L)$. 
\STATE Exchange $g_2[\,][\,]$, $h_2[\,][\,]$ and $g_1[\,][\,]$, $h_1[\,][\,]$.
\ENDWHILE
\STATE Rearrange $h_1[\,][\,]$ in the original order.
\STATE \RETURN $h_1[\,][\,]$
\end{algorithmic}
\end{algorithm}

\begin{algorithm}
\caption{Subprogram for the sliding sum in GPU (SSSG).}
  \label{al:ssGPU}
\begin{algorithmic}
\REQUIRE $g_1[\,][\,]$, $h_1[\,][\,]$, $L$ 
\STATE $s[x_{\rm T}][y_{\rm T}] \gets g_1[x_{\rm T} + 8 y_{\rm T} + 64 x_{\rm B}][y_{\rm B}]$.
\STATE $t[x_{\rm T}][y_{\rm T}] \gets h_1[x_{\rm T} + 8 y_{\rm T} + 64 x_{\rm B}][y_{\rm B}]$.
\STATE $r \gets 0$.
\WHILE{$r < 3$}
\IF{$x_{\rm T} < 16 - 2^r$}
\IF{$B(L,\,r) = 1$}
\STATE $t[x_{\rm T}][y_{\rm T}] \gets s[x_{\rm T}][y_{\rm T}] + t[x_{\rm T} + 2^r][y_{\rm T}]$.
\ENDIF
\STATE $s[x_{\rm T}][y_{\rm T}] \gets s[x_{\rm T}][y_{\rm T}] + s[x_{\rm T} + 2^r][y_{\rm T}]$.
\ENDIF
\STATE $r \gets r + 1$.
\ENDWHILE
\IF {$x_{\rm T} < 8$}
\STATE $g_2[x_{\rm T} + 8 x_{\rm B}][y_{\rm T} + 8y_{\rm B}] \gets s[y_{\rm T}][x_{\rm T}]$
\STATE $h_2[x_{\rm T} + 8 x_{\rm B}][y_{\rm T} + 8y_{\rm B}] \gets t[y_{\rm T}][x_{\rm T}]$
\ENDIF
\STATE \textbf{return} ($g_2[\,][\,]$, $h_2[\,][\,]$, $\lfloor L / 8 \rfloor$)
\end{algorithmic}
\end{algorithm}

Algorithm \ref{al:ssCPU} shows the main program for the sliding sum that works in a CPU. 
Algorithm \ref{al:ssGPU} shows the subprogram for the sliding sum that works in a GPU. 
Fig.~\ref{fig:slidingsumS} illustrates the calculation of $s[\,][\,]$ in a block
In this and the next figures, the second indexes are abbreviated. 
Fig.~\ref{fig:slidingsumT} illustrates the calculation of $t[\,][\,]$
in a part of a block when $B(L,\,0) = 1$, $B(L,\,1) = 0$, and $B(L,\,2) = 1$.

The two sentences in the last 'if' sentence of the procedure SSSG are the same as
\begin{align*}
g_2[y_{\rm T} + 8 x_{\rm B}][x_{\rm T} + 8y_{\rm B}] & \gets s[x_{\rm T}][y_{\rm T}],\\
h_2[y_{\rm T} + 8 x_{\rm B}][x_{\rm T} + 8y_{\rm B}] & \gets t[x_{\rm T}][y_{\rm T}].
\end{align*}
In order to access the global memory continuously,
$x_{\rm T}$ and $y_{\rm T}$ are exchanged.

\begin{figure}
  \centering
  \includegraphics[keepaspectratio,width=7.4cm]{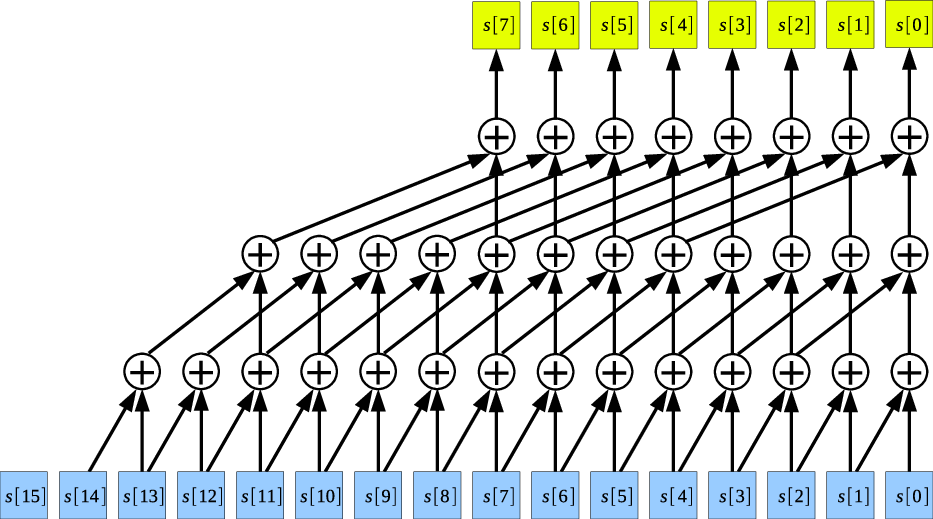}

  \
  
  \caption{Calculation of $s[\,][\,]$ for sliding sum in a block.
  (Its second index is abbreviated. )}
  \label{fig:slidingsumS}
\end{figure}

\begin{figure}
  \centering
  \includegraphics[keepaspectratio,width=7.4cm]{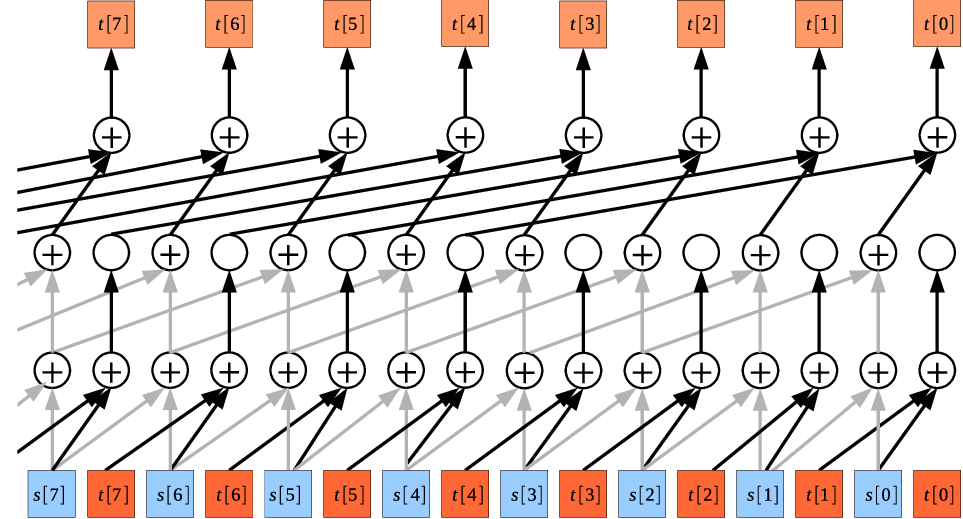}

  \
  
  \caption{Calculation of $t[\,][\,]$ for sliding sum in a part of block
    when $B(L,\,0) = 1$, $B(L,\,1) = 0$, and $B(L,\,2) = 1$.
      (Its second index is abbreviated. )}
  \label{fig:slidingsumT}
\end{figure}

When we calculate SFT by this algorithm of calculating the process
for each $p$ (the order of sinusoidal function) in the individual core,  
the calculation time can be reduced to $O((\log_2 P) (\log_2 K))$,
where $P$ is the total number of orders in SFT.
However, the number of cores for the calculation becomes $2PN$,
although it is $2N$ when calculations for all $p$ are done in a core.
The algorithm with $P = P_{\rm D} = 6$ has sufficient precision as shown in Section V so that
$P$ and $\log_2 P$ are not very different.
Furthermore, because the algorithm becomes complicated,
we use an algorithm where the calculations for all $p$ are done in a core.

When we use recursive filters,
values of all data whose length is $N$ should be added or subtracted with weights
to obtain a result at the last point.
Therefore, the floating point errors of single-precision floating point calculations
should be considered.
For this purpose, ASFT was proposed.
However, by this implementation, only values in the interval of length $2K + 1$
are summed to obtain a result at every point;
we do not have to consider floating point error and
can use SFT even with single-precision floating point calculations.

\begin{table}[htbp]
\caption{Abbreviations of filters.}
\begin{center}
\begin{tabular}{|l|c|c|c|c|c|}
  \hline
Abbreviation & Transform & Method & $P_{\rm D}$ or $P_{\rm M}$ & SFT/ASFT \\
  \hline
GDP6 & Gaussian & Direct  & 6 & SFT \\
MDP5 & Morlet & Direct  & 5 & SFT \\
MDP6 & Morlet & Direct  & 6 & SFT \\
MDP7 & Morlet & Direct  & 7 & SFT \\
MDP9 & Morlet & Direct  & 9 & SFT \\
MDP11 & Morlet & Direct  & 11 & SFT \\
MDS5P5 & Morlet & Direct & 5 & ASFT \\
MDS5P7 & Morlet & Direct & 7 & ASFT \\
MDS5P9 & Morlet & Direct  & 9 & ASFT \\
MDS5P11 & Morlet & Direct & 11 & ASFT \\
MMP2 & Morlet & Multiply  & 2 & SFT \\
MMP3 & Morlet & Multiply  & 3 & SFT \\
MMP4 & Morlet & Multiply  & 4 & SFT \\
MMP5 & Morlet & Multiply  & 5 & SFT \\
MMS5P2 & Morlet & Multiply  & 2 & ASFT \\
MMS5P3 & Morlet & Multiply  & 3 & ASFT \\
MMS5P4 & Morlet & Multiply  & 4 & ASFT \\
MMS5P5 & Morlet & Multiply  & 5 & ASFT \\
  \hline
\end{tabular}
\vspace{2mm}

`GCT3' and `MCT3' are the Gaussian smoothing and the Morlet wavelet transform realized by convolution with the function truncated
in interval $[-3 \sigma, \, 3 \sigma]$.
\label{nameOfFilter}
\end{center}
\end{table}

\begin{figure*}

  \begin{center}
\includegraphics[width=7.5cm]{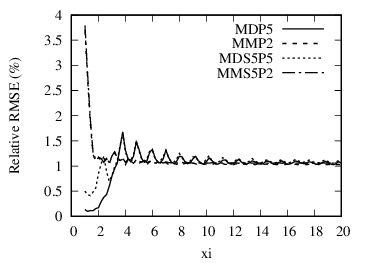}
\includegraphics[width=7.5cm]{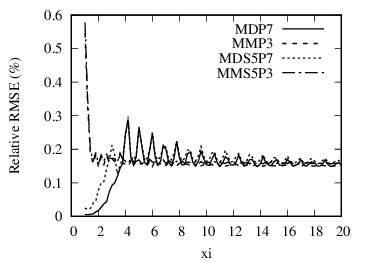} 
  \end{center}

\hspace{3cm}(a) $P_{D}=5$ and $P_{M}=2$
\hspace{4cm}(b) $P_{D}=7$ and $P_{M}=3$ 

\

  \begin{center}
\includegraphics[width=7.5cm]{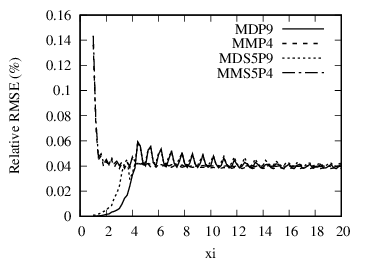}
\includegraphics[width=7.5cm]{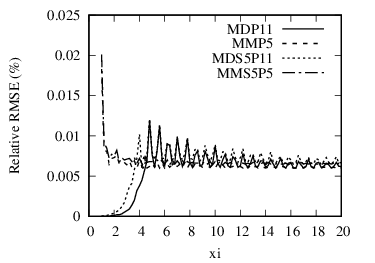} 
  \end{center}

\hspace{3cm}(c) $P_{D}=9$ and $P_{M}=4$
\hspace{4cm}(d) $P_{D}=11$ and $P_{M}=5$  
\caption{Relative RMSE of the Morlet wavelet function by SFT and ASFT.}
\label{fig:MSE1}
\end{figure*}

\begin{figure}
  \centering

\includegraphics[width=8.45cm]{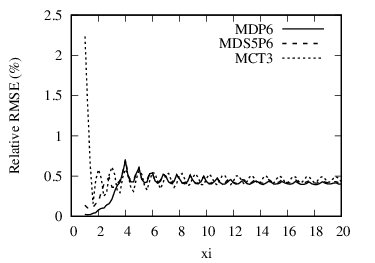} 

\caption{Relative RMSE of the Morlet wavelet function by SFT and ASFT ($P_{\rm D} =6$) and
  truncated one ($[-3 \sigma, 3 \sigma]$).}
\end{figure}

\begin{figure}
  \centering

\includegraphics[width=8.45cm]{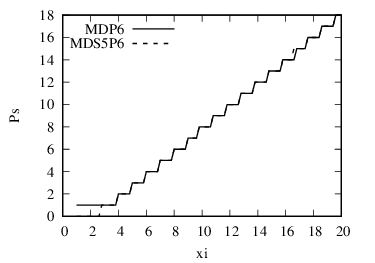}

\caption{Optimum $P_S$ to approximate the Morlet wavelet function by SFT ($P_{\rm D} =6$).}
\label{fig:MSE2}
\end{figure}

\section{Experiments}
The purpose of experiments is twofold.
The first one is to investigate errors to approximate the Morlet wavelet function.
By this experiment,
we can decide whether the direct method or the multiplication method should
be used and determine an optimum $P_{\rm S}$ for the direct method.
The second one is to evaluate calculation times of the proposed calculation algorithm for GPU.

Table \ref{nameOfFilter} describes the abbreviation of each of transforms and algorithms in the plots.
\begin{figure*}
  \begin{center}
\includegraphics[width=7.5cm]{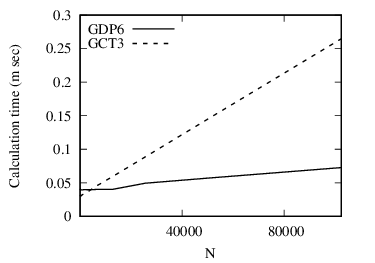}
\includegraphics[width=7.5cm]{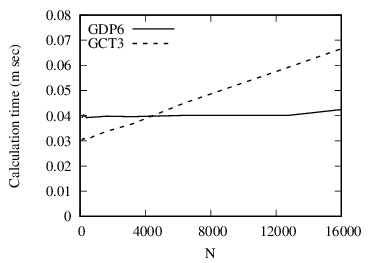}
  \end{center}
\hspace{1cm} (a) Calculation time versus to $N$ ($\sigma = 16$)  \hspace{2cm}
(b) Magnified horizontal scale in (a)

\

  \begin{center}
\includegraphics[width=7.5cm]{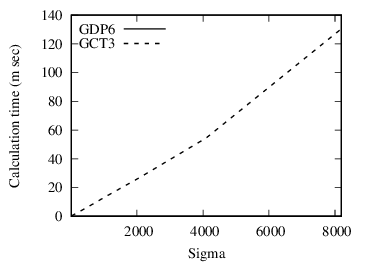}
\includegraphics[width=7.5cm]{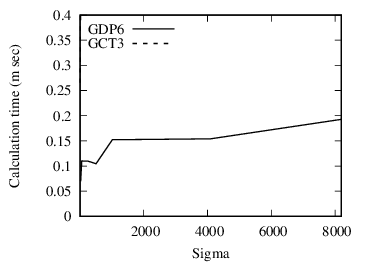}
  \end{center}
\hspace{1cm}(c) Calculation time versus $\sigma$ ($N = 102400$)  \hspace{2cm}
(d) Magnified vertical scale in (c)

\caption{Calculation time of Gaussian smoothing by SFT and truncated convolution using GPU.}
\label{fig:timeG}
\end{figure*}

\begin{figure*}
  \begin{center}
\includegraphics[width=7.5cm]{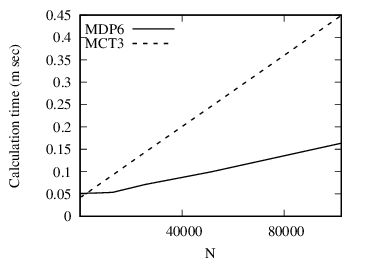}
\includegraphics[width=7.5cm]{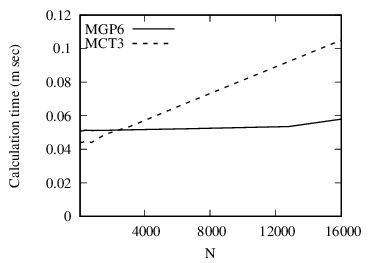}
  \end{center}
\hspace{1.5cm}(a) Calculation time versus to $N$ ($\sigma = 16$)  \hspace{2.0cm}
(b) Magnified horizontal scale in (a)

\

  \begin{center}
\includegraphics[width=7.5cm]{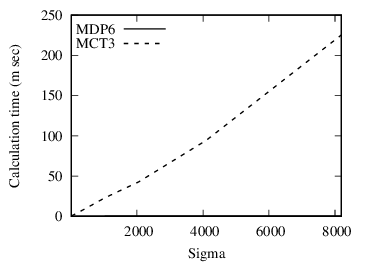}
\includegraphics[width=7.5cm]{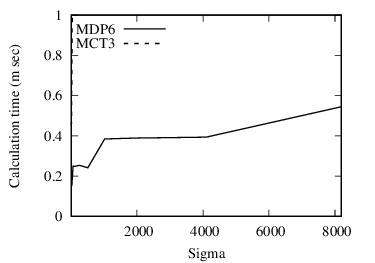}
  \end{center}
\hspace{1.2cm}(c) Calculation time versus $\sigma$ ($N = 102400$)  \hspace{2.0cm}
(d) Magnified vertical scale in (c)

\caption{Calculation time of the Morlet wavelet transform by SFT and truncated convolution using GPU.}
\label{fig:timeM}
\end{figure*}

\subsection{Approximation error of the Morlet wavelet}

Let $\hat{\psi}_{\sigma, \,\xi}[n]$ be an approximated Morlet wavelet function.
The relative RMSE is given by
\begin{eqnarray}
\mbox{Relative RMSE} = \sqrt{\frac{\sum_{n=-5K}^{5K}|\hat{\psi}_{\sigma, \,\xi}[n] - \psi_{\sigma, \,\xi}[n]|^2}{\sum_{n=-5K}^{5K} |\psi_{\sigma, \,\xi}[n]|^2} }. 
\end{eqnarray}
Because $K$ is around $3 \sigma$, the interval of summation is sufficient to evaluate the error.

In this experiment, we fix $\sigma = 60.0$ and change $\xi$ from 1.0 to 20.0.
$K$ is chosen such that the relative RMSE becomes the smallest.

Fig.~\ref{fig:MSE1} shows the relative RMSEs of the approximated Morlet wavelet function using SFT and ASFT with various $P_D$ and $P_M$
for the direct and multiply methods.
We can observe that if $P_D = 2 P_M + 1$, their relative RMSEs are almost the same for 
$\xi \geq 6$.
However, when $\xi$ is small, the relative RMSEs of the multiply method is larger
than those of the direct method.
When we calculate the Morlet wavelet transform,
we must calculate the SFT/ASFT of $P_D$ and $3P_M + 1$ orders.
Furthermore, because the wavelength of sinusoidal functions for the direct method
is $K$ divided by an integer, the calculation of filters can be simplified.
However, because those for the multiplication method are real numbers in general,
it cannot be simplified. 
Therefore, it would be better  to use the direct method for the Morlet wavelet transform 
although we have to select $P_{\rm S}$ according to $\xi$.

There is minimal difference between SFT and ASFT.
Therefore, we can use ASFT when $N$ is large;
moreover, we use recursive filers with single-precision floating point calculations.

Fig.~\ref{fig:MSE1} (a) shows relative RMSEs of the Morlet wavelet function
approximated by the direct method with SFT and ASFT of $P_{\rm D} =6$ and
that of the Morlet wavelet function truncated in interval the $[-3 \sigma, 3 \sigma]$.
We can see that relative RMSEs of both methods are almost the same.
Therefore, we will compare the calculation speeds
between them.
Fig.~\ref{fig:MSE2} (b) shows the optimum $P_{\rm S}$.
We find that it increases as $\xi$ increases. 

\subsection{Calculation speed}

We compare the calculation time of the proposed method and
that of the convolution between an input signal and a transform function truncated by the interval $[-3 \sigma, 3 \sigma]$.
We denote the latter method as truncated convolution. 
When the truncated convolution is calculated by GPU with the parallel reduction \cite{ParallelReduction201X},
the calculation time of the summation in the interval is approximately reduced to $\log_2 (6 \sigma + 1)$.
We use a computer comprising
AMD Ryzen 9 3900X (12 cores, 3.8 GHz) and RTX 3090 (10496 CUDA COREs, 1.70 GHz).

Fig.~\ref{fig:timeG} and Fig.~\ref{fig:timeM}
show calculation times for Gauss smoothing and the Morlet wavelet transform, respectively.
Fig.~\ref{fig:timeG} (a) and Fig.~\ref{fig:timeM} (a)
show the calculation times when we change $N$ from 100 to 102400 
while $\sigma $ is fixed at 16.0.
Fig.~\ref{fig:timeG} (b) and Fig.~\ref{fig:timeM} (b)
are those with magnified horizontal scale.
Fig.~\ref{fig:timeG} (c) and Fig.~\ref{fig:timeM} (c)
show the calculation times when we change $\sigma$ from 16.0 to 8192.0 
while $N$ is fixed at 102400.
Fig.~\ref{fig:timeG} (d) and Fig.~\ref{fig:timeM} (d)
are those with magnified vertical scale.

In Fig.~\ref{fig:timeG} (c), Fig.~\ref{fig:timeG} (d), Fig.~\ref{fig:timeM} (c), and Fig.~\ref{fig:timeM} (d), 
the graphs for the truncated convolution can be seen only in (c) and
those of the proposed method can be seen only in (d)
because the calculation times are different. 

Let $M$ be the number of cores. 
In this experiment, we use $M = 10496$.
In the truncated convolution, 
the number of multiplications is approximately $N(6\sigma + 1)$.
If $M$ is larger than this, the order of its calculation time is one.
Otherwise, it is $O(N \sigma /M)$.
The number of additions is also approximately $N (6\sigma + 1)$.
If $M$ is larger than this, the order of its calculation time is almost $O(\log_2 \sigma)$.
Otherwise, it is $O((N \log_2 \sigma) / M)$.

In the proposed method, 
the number of multiplications is approximately $7NP$.
If $M$ is larger than $N$, the order of its calculation time is $O(P)$.
(As discussed in Section IV, although it can be reduced to $O(1)$ 
using a core for each data point and each order of SFT ($p$).
The procedure, however, becomes complicated to integrate their results.) 
If $M$ is not larger than $N$, it is $O(NP/ M)$.
The number of additions is also approximately $NP(2K + 1)$.
If $M$ is larger than $N$, the order of calculation time is  $O(P\log_2 K)$.
Otherwise, it $O(N P(\log_2 K) / M)$.

From Figs. \ref{fig:timeG} (b) and \ref{fig:timeM} (b),
it is found that if $N$ or $\sigma$ is large,
the proposed method has a clear advantage in the calculation time.
Only when both $N$ and $\sigma$ are small,
is the truncated convolution a little faster than the proposed method.
In such cases, although the order of calculation time is almost the same,
the algorithm of the truncated convolution is simpler
and hence, a little faster.
However, especially when $\sigma$ is large,
we can see that the proposed method is much faster
because the number of multiplications of the proposed method is much smaller
than that of the truncated convolution.

The order of memory accesses for the truncated convolution and the proposed method are approximately $O(N \sigma \log_2 \sigma)$ and $O(N P \log_2 K)$.
Because we use the global memory as well as shared memories, its evaluation is difficult. 
However, 
we know that the proposed method is faster for a large $\sigma$
because $P$ does not depend on $\sigma$ and the orders of $\sigma$ and $K$ are almost the same.

\section{Conclusion}
Based on the calculation methods for Gaussian smoothing using SFT/ASFT, 
we proposed two types of calculation methods for the Morlet wavelet transform. 
We also proposed an algorithm to calculate SFT using the kernel integral on GPU.
When the number of calculation cores in GPU is not less than the number of data points,
the order of the calculation time is proportional to the logarithm of the smoothing size
(standard deviation of the Gaussian function) 
and does not depend on the number of data points.
Experimental results successfully showed the properties of the proposed methods and 
the advantages of the proposed algorithm.
For future work, it is important to apply the methods to real-world problems, 
and, also,  apply the algorithm to other transformations.



\end{document}